\documentclass{article}
\usepackage{hiph-art}
\volnumber{ } \issuenumber{} \edyear{2005}                             
\frompage{000} \topage{000}                                              
\recrevdate{29 October 2005}                                              
\def\rightmark{Inclusive production of $\Lambda$, $K^0_s$ and narrow resonances }
\def\leftmark{Aslanyan}

\title{Inclusive production of $\Lambda$, $K^0_s$ and
exotic narrow resonances for systems $K_s^0 p$, $K_s^0 \Lambda$,
$\Lambda p$ from p+propane interactions at 10 GeV/c  }
\authors{{P.Zh.Aslanyan$^{1,2}$
\index{Aslanyan} 
 }\\[2.812mm]
 {\normalsize
\hspace*{-8pt}$^1$ Joint Institute for Nuclear Research,VBLHE\\[0.2ex]
\hspace*{-8pt}$^2$ Yerevan State University }}
\abstract{Experimental data from the 2m propane bubble chamber for
production of $\Lambda$, $K^0_s$ have been used to search of exotic
baryon states, in the $K_s^0 p$, $K_s^0 \Lambda$ and $\Lambda p$
decay mode for the reaction p+propane at 10 GeV/c. The estimation of
experimental inclusive cross sections for $\Lambda$ and $K^0_s$
production in the p$^{12}C$ collision is equal to
$\sigma_{\Lambda}$= 13.3$\pm$1.7 mb and $\sigma_{K^0_s}$=
3.8$\pm$0.6 mb, respectively. The measured $\Lambda /\pi^+$ ratio
from pC reaction is equal to (5.3$\pm0.8)*10^{-2}$. The experimental
$\Lambda /\pi^+$ ratio in the pC reaction is approximately two times
larger than the $\Lambda /\pi^+$ ratio simulated by FRITIOF model in
the pC reaction.
 The invariant mass spectrum $\Lambda K^0_s$ registered
 narrow peaks in regions of 1750 and 1795 MeV/$c^2$. The
statistical significance of these peaks has been estimated as 5.6
and 3.3 S.D., respectively. These would be candidates for the $N^0$
or the $\Xi^0$ pentaquark states.
 The $pK^0_s$ invariant mass spectrum shows resonant structures with
$M_{K_s^0 p}$=1540, 1613, 1821 MeV/$c^2$. The statistical
significance of these peaks have been estimated as 5.5,4.8 and 5.0
s.d., respectively.
   The invariant mass spectrum S=-1 $\Lambda p$ observed a narrow peaks
at 2100, 2175,2285 and 2353 MeV/$c^2$. Their excess above background
by the second method is 6.9, 4.9, 3.8 and 2.9 S.D., respectively.
 }

 \makeindex
\begin{document}
  \maketitle
\footnote{Permanent address: JINR,VBLHE, Dubna, Moscow region,
Joliot Curie 6, Russia; E-mail: paslanian@jinr.ru}

\section{Introduction}\label{intro}

 Strange particle production \cite{1} has been analyzed regarding such
 reaction mechanisms as the multinucleon effect, or the fireball effect,
 or possible signature for the quark-gluon plasma(QGP), or as the deconfiment
signal, within the context of thermal equilibration models. In
particular, the $\Lambda /\pi^+$ ratio of particle production  have
been observed extensively on hadron - nucleus and nucleus-nucleus
collisions 4-15 GeV regions.

Multi-quark states, glueballs and hybrids have been searched for
experimentally for a very long time, but none is established.
Several models (\cite{2}-\cite{5}) predict the multiplet structure
and characteristics of multi quark hadrons and pentaquarks.

 Results from a wide range of recent experiments\cite{2} are
 consistent with the  existence of an exotic S=+1 resonance,
 the $\Theta^+(1540)$ with a narrow width and a mass near 1540 MeV.

Preliminary results on a search for the  $N^0$ or the $\Xi^0$
pentaquark states in the decay mode $\Lambda K^0_s$ with the mass
$1734\pm0.5\pm5$ MeV/$c^2$  is presented in the article S. Kabana,
hep-ph/0501121, 2005.

 Metastable strange  dibaryons were searched a long time ago at LHE JINR, too.
  This group succeeded in finding resonance-like peaks
\cite{3,4} only in five of them $\Lambda p$, $\Lambda p \pi$,
$\Lambda \Lambda$, $\Lambda \Lambda p $, $\Lambda \pi^+ \pi^+$.

\section{Experimental procedure}\label{exp}

\subsection{Identification of $\Lambda$ and $K^0_s$}\label{v0}
The experimental events with $V^0$s were searched on $\approx$
700000 stereo photographs from the 2m propane bubble chamber of
JINR, LHE exposed proton beams at 10 GeV/c \cite{1}.
 The events with $V^0$ ($\Lambda$ and $K^0_s$)  were identified
using the following criteria: 1) $V^0$ stars from the photographs
were selected according to $\Lambda\to\pi^-+p$, neutral
$K_s\to\pi^-+\pi^+$ or $\gamma \to e^++e^-$ hypothesis. A momentum
limit of $K^0_s$ and $\Lambda$ is greater than 0.1 and 0.2 GeV/c,
respectively ; 2) $V^0$ stars should have the effective mass of
$K^0_s$ and of $\Lambda$; 3) these $V^0$ stars are directed to  some
vertices(complanarity); 4) they should have one vertex, a three
constraint fit for the $M_K$ or $M_{\Lambda}$ hypothesis  and after
the fit, $\chi^2_{V^0}$ should be selected  over range less than 12;
5)The analysis has shown  that the events with undivided $\Lambda
K^0_s$ were assumed to be events as $\Lambda$(Fig.1a,1b).  As a
result of above procedure have lost of $K^0_s$ 8.5\% and admixture
of $K^0_s$ in $\Lambda$s events 4.6\%.

Fig. 1c and 1d shows the effective mass distribution of
$\Lambda$(8657-events) and $K^0$(4122-events) particles,
respectively. The masses of the observed $\Lambda$, $K^0_s$ are
consistent with their PDG values.

\section{The measured cross sections $\Lambda$ and $K^0$}\label{cross}

The cross section is defined by the formula:
$$
  \sigma = \frac{\sigma_0*N_r^{V^0}}{e}\prod_iw_i=
\frac{\sigma_r*N_r^{V^0}*w_{hyp}*w_{geom}*w{\phi}*w_{kin}*w_{int}}{N_r*e_1*e_2*e_3},
(3.1)
$$
 where  $e_1$ is the efficiency of search for $V^0$ on the photographs,
$e_2$ the efficiency of measurements. The $V^0$s of
75\%(preliminary) could be successfully reconstructed and accepted
in the analysis. $e_3$ the probability of decay via the  channel of
charged particles ($\Lambda\to p\pi^-, K^0\to\pi^+\pi^-$),
$\sigma_0= \sigma_r/N_r$ the total cross section, where $\sigma_r$
and $N_ r $ is the total cross section and number of registered
events. The propane bubble chamber method have been permitted the
registration of the part of all elastic interactions with the
propane \cite{1}, therefore the total cross section of registered
events is equal to:$\sigma_r(p+C_3H_8)=3\sigma_{pC}(inelastic)+
8\sigma_{pp}(inelastic)+8\sigma_{pp}(elastic)0.70=(1049\pm60)$mb.
$w_i$ are  weights for the lost events with $V^0$ for: $w_{geom}$ -
the $V^0$ decay outside the chamber; $w_{\phi}$ - the required
isotropy for $V^0$ in the azimuthal (XZ) plane; $w_{hyp}$ - the
undivided $\Lambda K^0_s$ events; $w_{int}$ - the selected as
$p~+~^{12}C$ from the interaction of $p~+~C_3H_8$; $w_{kin}$ - the
kinematic conditions(with FRITIOF);$w_{int}$- the $V^0$+ propane
interactions. The criteria for selection of interaction with carbon
has shown\cite{1}.

Table~\ref{tab:evcros} show that the experimental cross sections are
calculated by formula 3.1  for inclusive productions $\Lambda$
hyperons and $K_s^0$ mesons  for the interactions of pp and pC at
beam momentum 10 GeV/c. The experimental data of multiplicities
$pi^+$ mesons taken from experiments. The experimental $\Lambda
/\pi^+$ ratio from the pC reaction is approximately two times larger
than the experimental $\Lambda /\pi^+$ ratio  for pp  and
 the simulated FRITIOF model from pC reactions. The $\Lambda /\pi^+$
  ratio for C+C reaction at momentum 10 GeV/c have been obtained by using the
Glauber approach on the experimental cross section for p+C
$\to\Lambda $X reaction(Fig.2).  As can be seen from experimental
data  and thermal statistical model (Fig.2) there is a very clearly
pronounced enhancement specially in the $\Lambda /\pi^+$ ratio for
hadron-nucleus and nucleus collisions at 10-15 A.GeV/c.

\section{$pK^0_s$,$\Lambda K^0_s$ and $\Lambda p$ spectrum  analysis}

\subsection{The experimental background}

The total experimental background has been  obtained by three
methods \cite{2},\cite{5}. In the first method, the experimental
effective mass distribution was approximated by the polynomial
function  after cutting out the resonance ranges because this
procedure has to provide the fit with $\Xi^2$=1 and polynomial
coefficient with errors less than 30 \%.  The second of the randomly
mixing method of the angle between of  decaying particles from the
resonance for experimental events is described in V.L.Lyuboshits et
al.(see \cite{2}).  The third background method has been obtained by
using FRITIOF model  with experimental conditions. The analysis done
by three methods has shown that while fitting these distributions
had the same coefficients and order of the polynomial function. The
values for the mean position of the peak and the width obtained by
using Breit Wigner fits.

\subsection{$pK^0_s$ - spectrum for protons with a momentum of
$0.350\le p_p\le 0.900$ GeV/c}

  The $pK^0_s$ effective mass distribution 2300 combination (Fig.3a) is
shown resonant structures with $M_{K_s^0 p}$=1540, 1613, 1821
MeV/$c^2$ and $\Gamma_{K_s^0 p}$= 9.2, 16.1, 28.0
MeV/$c^2$(\cite{2}). The statistical significance of these peaks
have been estimated as 5.5,4.8 and 5.0 s.d., respectively. There are
also small peaks in 1690( 3.6 s.d.), 1750 (2.3 s.d.) and 1980(3.0
s.d.) MeV/$c^2$ mass regions.

\subsection{$\Lambda K^0_s$ - spectrum  analysis}

Figure 3b  shows  the invariant mass of 1012 ($\Lambda
K^0_s$)combinations with bin sizes 18 MeV/$c^2 $(\cite{5}). There
are significant enhancements in mass regions of 1750 and 1795
MeV/$c^2$(Fig.3b). Their excess above background by the first method
is 5.0 and 3.0 S.D,respectively. There are small enhancement in mass
regions of 1670,1850 and 1935 MeV/$c^2$.

\subsection{$\Lambda p$ - spectrum  analysis for protons with a momentum of
$0.250\le p_p\le 0.900$ GeV/c}

Figure  3c  shows  the invariant mass of 2434 ($\Lambda
p$)combinations with bin sizes 15 MeV/$c^2 $(\cite{4}). The values
for the mean position of the peak and the width obtained by using
Breit Wigner fits. There are significant enhancements in mass
regions of 2100, 2175, 2285 and 2353 MeV/$c^2$(Fig.1c).Their excess
above background by the second method is 6.9, 4.9, 3.8 and 2.9 S.D.,
respectively. There is also a small peak in 2225( 2.2 s.d.)
MeV/$c^2$ mass region.

\section{Conclusion}

The experimental data from the 2 m propane bubble chamber  have been
analyzed from pC$\to\Lambda(K^0_s)$X reactions at 10 GeV/c. The
estimation of experimental inclusive cross sections of $\Lambda$ and
$K^0_s$ production for pC collisions is equal to $\sigma_{\Lambda}$=
13.3$\pm$1.7 mb and $\sigma_{K^0_s}$= 3.8$\pm$0.6 mb, respectively.
The measured $\Lambda /\pi^+$ ratio for pC is equal to (5.3 ±
0.8)*10-2.  The $\Lambda /\pi^+$ ratio for C+C collisions at 10 A
GeV/c obtained that is $\approx$ 3-4 times larger than the $\Lambda
/\pi^+$ ratio from C+C reactions at the same energy simulated by
FRITIOF model.

 A number of peculiarities were found in the effective mass
spectrum of $K^0_s p$, $\Lambda K^0_s$ and $\Lambda p$(section 4).
The $N^0$ can be  from  the antidecuplet, from an octet (D.
Diakonov, V. Petrov , V.Guzey and M.Polyakov or an 27-plet(J. Ellis
et al.). On the other hand, Jafe and Wilczek predicted a mass around
1750 MeV and a width 50 \% larger for these states than that of the
$\Theta^+$. These peaks in the effective mass spectrum $\Lambda
K^0_s$ are possible candidates for two pentaquark states: the $N^0$
with quark content udsds decaying into $\Lambda K^0$ and the $\Xi^0$
quark content udssd decaying into $\Lambda \overline{K^0}$. The
calculated rotational $\Theta+$ and $N^0$ spectra from theoretical
reports of D.Akers, Y Nambu, M.H. Mac Gregor and A.A. Arkhipov
agreed with experimental reports of Yu. A. Troyan and P.Zh Aslanyan
\cite{2} .

 The experimental result for S=-1  $\Lambda p$  dibaryon spectrum shows
that the predicted peaks with the bag model has been
confirmed\cite{3}.

\begin{figure}[htb]
\vspace*{0.1cm} \epsfysize=60mm
 \centerline{
 \epsfbox{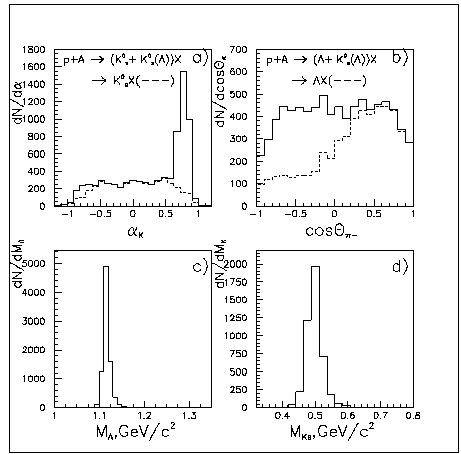}}
\vspace*{0.cm} \caption[]{(a) and (b) distributions  of  $\alpha$
(Armenteros parameter) and  cos$\Theta^*$-
  are used  for  correctly identification  of   the undivided
  $V^0$s. $\alpha = (P^+_{\parallel}-P^-_{\parallel})/((P^+_{\parallel}+P^-_{\parallel})$. Where $P^+_{\parallel}$
  and  $P^-_{\parallel}$ are the parallel components of momenta  positive and  negative
  charged  tracks. cos$\Theta^*$ - is the angular distribution  of $\pi^-$ from
  $K_s^0$  decay. Distributions of $\alpha$  and  cos$\theta$- were isotropic in the rest  frame
  of $K_s^0$  when  undivided $\Lambda K_s^0$ were assumed to be  events  as $\Lambda$.
  c) and (d) distributions of experimental $V^0$ events produced
 from interactions of beam protons with propane: c) for the effective mass of
  $M_{\Lambda}$; d) for the effective mass of $M_{K^0_s}$.
  }
\label{fig1}
\end{figure}

 \begin{figure}[htb]
\vspace*{-0.3cm} \epsfysize=70mm
 \centerline{
 \epsfbox{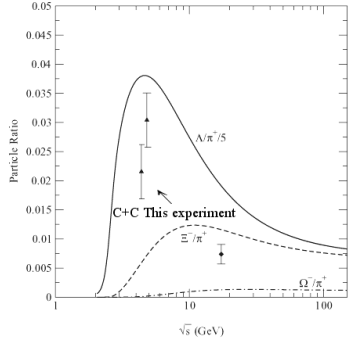}}
\vspace*{-2.5cm} \caption[]{Prediction of the statistical-thermal
model(note the factor 5) for $\Lambda/\pi^+$(solid line), and
$\Xi^-/\pi^+$(dashed line) and $\Omega^-/\pi^+$ ratios a function of
$\sqrt{s}$. For compilation of AGS data see(above point)\cite{1}.
The $\Lambda/\pi^+$ ratio is presented by using data from this
experiment for C+C interaction. } \label{fig2}
\end{figure}

\begin{figure}[htb]
\vspace*{-0.3cm}
\epsfysize=40mm \epsfxsize=150mm
 \centerline{
 \epsfbox{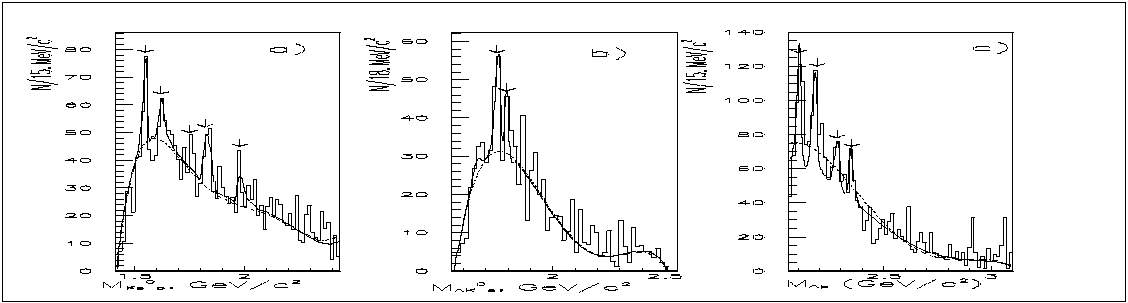}}
\vspace*{-0.0cm} \caption[]{The effective mass distribution for
systems:(a) $pK^0_s$,(b)$\Lambda K^0_s$ and (c)$\Lambda p$.}
\label{albw}
\end{figure}

\begin{table}
\begin{tabular}{lrlrlrlrlr}
\hline
Type of &$N_{V^0}^{exp.}$&$W_{sum}$&$N_{V^0}^t$&$n_{V^0}=N_{V^0}^t/N_{in}$&$\sigma$ \\
reaction& &&Total&&mb\\ \hline pC$\to
\Lambda X$&6126&4.37$\pm$0.37&26770&0.053$\pm$0.005&13.3$\pm1.6$\\
\hline pp$\to \Lambda
X$&836&5.15$\pm$0.44&4303&0.026$\pm$0.003&0.80$\pm$0.08
\\\hline pC$\to K^0_s
X$&3188&2.93$\pm$0.25&9341&0.018$\pm$0.002&3.8$\pm$0.5
\\\hline
pp$\to K^0_s
X$&699&3.31$\pm$0.28&2313&0.015$\pm$0.001&0.43$\pm$0.04\\\hline
\end{tabular}
\caption{ Cross sections  $\Lambda$ hyperons and $K_s^0$ mesons for
pp and pC interactions at  beam momentum 10 GeV/c.}
 \label{tab:evcros}
\end{table}

\vfill\eject
\end{document}